# Empirical Analysis and Statistical Modeling of Attack Processes based on Honeypots

M. Kaâniche[1], E. Alata[1], V. Nicomette[1], Y. Deswarte[1], M. Dacier[2]

[1]*LAAS-CNRS, Université de Toulouse*
*7 Avenue du Colonel Roche, 31077 Toulouse Cedex 4, France*
*{kaaniche, ealata, deswarte, nicomett}@laas.fr*

[2]Eurécom
*2229 Route des Crêtes, BP 193, 06904 Sophia Antipolis Cedex, France*
*dacier@eurecom.fr*

**Abstract**

*Honeypots are more and more used to collect data on malicious activities on the Internet and to better understand the strategies and techniques used by attackers to compromise target systems. Analysis and modeling methodologies are needed to support the characterization of attack processes based on the data collected from the honeypots. This paper presents some empirical analyses based on the data collected from the Leurré.com honeypot platforms deployed on the Internet and presents some preliminary modeling studies aimed at fulfilling such objectives.*

## 1. Introduction

Several initiatives have been developed during the last decade to monitor malicious threats and activities on the Internet, including viruses, worms, denial of service attacks, etc. Among them, we can mention the Internet Motion Sensor project [1], CAIDA [2], DShield [3], and CADHo [4]. These projects provide valuable information on security threats and the potential damage that they might cause to Internet users. Analysis and modeling methodologies are necessary to extract the most relevant information from the large set of data collected from such monitoring activities that can be useful for system security administrators and designers to support decision making. The designers are mainly interested in having representative and realistic assumptions about the kind of threats and vulnerabilities that their system will have to cope with once it is used in operation. Knowing who are the enemies and how they proceed to defeat the security of target systems is an important step to be able to build systems that can be resilient with respect to the corresponding threats. From the system security administrators' perspective, the collected data should be used to support the development of efficient early warning and intrusion detection systems that will enable them to better react to the attacks targeting their systems.

As of today, there is still a lack of methodologies and significant results to fulfill the objectives described above, although some progress has been achieved recently in this field. The CADHo project "Collection and Analysis of Data from Honeypots" [4], an ongoing research action started in September 2004, is aimed at contributing to filling such a gap by carrying out the following activities:

1) deploying a distributed platform of honeypots [5] that gathers data suitable to analyze the attack processes targeting a large number of machines connected to the Internet;
2) developing analysis methodologies and modeling approaches to validate the usefulness of this platform by carrying out various analyses, based on the collected data, to characterize the observed attacks and model their impact on security.

A honeypot is a machine connected to a network but that no one is supposed to use. In theory, no connection to or from that machine should be observed. If a connection occurs, it must be, at best an accidental error or, more likely, an attempt to attack the machine.

The *Leurré.com* data collection environment [5], set up in the context of the CADHo project, has deployed, as of to date, thirty five honeypot platforms at various locations from academia and industry, in twenty five countries over the five continents. Several analyses carried out based on the data collected so far from these honeypots have revealed that very interesting observations and conclusions can be derived with respect to the attack activities observed on the Internet [4, 6-9]. In addition, several automatic data analyses and clustering techniques have been developed to facilitate the extraction of relevant information from the collected data. A list of papers detailing the methodologies used and the results of these analyses is available in [6].

This paper focuses on modeling-related activities based on the data collected from the honeypots. We first discuss the objectives of such activities and the challenges that need to be addressed. Then we present some examples of models obtained from the data.

The paper is organized as follows. Section 2 presents the data collection environment. Section 3 focuses on the modeling of attacks based on the data collected from the honeypots deployed. Modeling examples are presented in Section 4. Finally, Section 5 discusses future work.

## 2. The data collection environment

The data collection environment (called *Leurré.com* [5]) deployed in the context of the CADHo project is based on low-interaction honeypots using the freely available software called *honeyd* [10]. Since September 2004, 35 honeypot platforms have been progressively deployed on the Internet at various geographical locations. Each platform emulates three computers running Linux RedHat, Windows 98 and Windows NT, respectively, and various services such as ftp, web, etc. A firewall ensures that connections cannot be initiated from the computers, only replies to external solicitations are allowed. All the honeypot platforms are centrally managed to ensure that they have exactly the same configuration. The data gathered by each platform are securely uploaded to a centralized database with the complete content, including payload of all packets sent to or from these honeypots, and additional information to facilitate its analysis, such as the IP geographical localization of packets' source addresses, the OS of the attacking machine, the local time of the source, etc.

## 3. Modeling objectives

Modeling involves three main steps:
1) The definition of the objectives of the modeling activities and the quantitative measures to be evaluated.
2) The development of one (or several) models that are suitable to achieve the specified objectives.
3) The processing of the models and the analysis of the results to support system design or operation activities.

The data collected from the honeypots can be processed in various ways to characterize the attack processes and perform predictive analyses. In particular, modeling activities can be used to:

- Identify the probability distributions that best characterize the occurrence of attacks and their propagation through the Internet.
- Analyze whether the data collected from different platforms exhibit similar or different malicious attack activities.
- Model the time relationships that may exist between attacks coming from different sources (or to different destinations).
- Predict the occurrence of new waves of attacks on a given platform based on the history of attacks observed on this platform as well as on the other platforms.

For the sake of illustration, we present in the following sections simple preliminary models based on the data collected from our honeypots that are aimed at fulfilling such objectives.

## 4. Examples

The examples presented in the following address:
1) The analysis of the time evolution of the number of attacks taking into account the geographic location of the attacking machine.
2) The characterization and statistical modeling of the times between attacks.
3) The analysis of the propagation of attacks throughout the honeypot platforms.

The data considered for the examples has been collected from January $1^{st}$, 2004 to April 17, 2005, corresponding to a data collection period of 320 days. We take into account the attacks observed on 14 honeypot platforms among those deployed so far. The selected honeypots correspond to those that have been active for almost the whole considered period. The total number of attacks observed on these honeypots is 816476. These attacks are not uniformly distributed among the platforms. In particular, the data collected from three platforms represent more than fifty percent of the total attack activity.

### 4.1 Attack occurrence and geographic distribution

The preliminary models presented in this sub-section address: i) the time-evolution modeling of the number of attacks observed on different honeypot platforms, and ii) the analysis of potential correlations for the attack processes observed on the different platforms taking into account the geographic location of the attacking machines and the proportion of attacks observed on each platform, wrt. to the global attack activity.

Let us denote by:
- *Y(t)* the function describing the evolution of the number of attacks per unit of time observed on all the honeypots during the observation period,

- $X_j(t)$ the function describing the evolution of the number of attacks per unit of time observed on all the honeypots during the observation period for which the IP address of the attacking machine is located in country $j$.

In a first stage, we have plotted, for various time periods, $Y(t)$ and the curves $X_j(t)$ corresponding to different countries $j$. Visual inspection showed surprising similarities between $Y(t)$ and some $X_j(t)$. To confirm such empirical observations, we have then decided to rigorously analyze this phenomenon using mathematical linear regression models.

Considering a linear regression model, we have investigated if $Y(t)$ can be estimated from the combination of the attacks described by $X_j(t)$, taking into account a limited number of countries $j$. Let us denote by $Y^*(t)$ the estimated model.

Formally, $Y^*(t)$ is defined as follows:

$$Y^*(t) = \Sigma \alpha_j X_j(t) + \beta \qquad j = 1, 2, .. k \qquad (1)$$

Constants $\alpha_j$ and $\beta$ correspond to the parameters of the linear model that provide the best fit with the observed data, and $k$ is the number of countries considered in the regression.

The quality of fit of the model is measured by the statistics $R^2$ defined by:

$$R^2 = \Sigma(Y^*(i) - Y_{av})^2 / \Sigma(Y(i) - Y_{av})^2 \qquad (2)$$

$Y(i)$ and $Y^*(i)$ correspond to the observed and estimated number of attacks for unit of time $i$, respectively. $Y_{av}$ is the average number of attacks per unit of time, taking into account the whole observation period.

Indeed, $R$ is the correlation factor between the estimated model and the observed values. The closer the $R^2$ value is to 1, the better the estimated model fits the collected data.

We have applied this model considering linear regressions involving one, two or more countries. Surprisingly, the results reveal that a good fit can be obtained by considering the attacks from one country only. For example, the models providing the best fit taking into account the total number of attacks from all the platforms are obtained by considering the attacks issued from either UK, USA, Russia or Germany only. The corresponding $R^2$ values are of the same order of magnitude (0.944 for UK, 0.939 for USA, 0.930 for Russia and 0.920 for Germany), denoting a very good fit of the estimated models to the collected data. For example, the estimated model obtained when considering the attacks from Russia only is defined by equation (3):

$$Y^*(t) = 44.568\, X_1(t) + 1555.67 \qquad (3)$$

$X_1(t)$ represents the evolution of the number of attacks from Russia. Figure 1 plots the evolution of the observed and estimated number of attacks per unit of time during the data collection period considered in this example. The unit of time corresponds to 4 days. It is noteworthy that, similar conclusions are obtained if we consider another granularity for the unit of time, for example one day, or one week.

These results are even more surprising that the attacks from Russia and UK represent only a small proportion of the total number of attacks (1.9% and 3.7% respectively). Concerning the USA, although the proportion is higher (about 18%), it is not sufficient to explain the linear model.

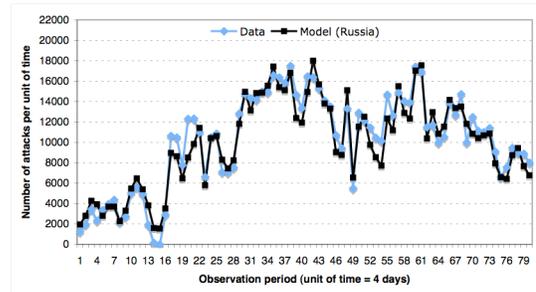

**Figure 1-** Evolution of the number of attacks per unit of time observed on all the platforms and estimated model considering attacks from Russia only

We have applied similar analyses by respectively considering each honeypot platform in order to investigate if similar conclusions can be derived by comparing their attack activities per source country to their global attack activities. The results are summarized in Table 1. The second column identifies the source country that provides the best fit. The corresponding $R^2$ value is given in the third column. Finally, the last three columns give the $R^2$ values obtained when considering UK, USA, or Russia in the regression model.

It can be noticed that the quality of the regressions measured when considering attacks from Russia only is generally low for all platforms ($R^2$ less than 0.80). This indicates that the property observed at the global level is not visible when looking at the local activities observed on each platform. However, for the majority of the platforms, the best regression models often involve one of the three following countries: USA, Germany or UK, which also provide the best regressions when analyzing the global attack activity considering all the platforms together. Two exceptions are found with P6 and P8 for which the observed attack activities exhibit different characteristics with respect to the origin of the attacks (Taiwan, China), compared to the other platforms.

The trends discussed above have been also observed when considering a different granularity for the unit of time (e.g., 1 day or 1 week) as well as different data observation periods.

| Platform | Country providing the best model | $R^2$ Best model | $R^2$ UK | $R^2$ USA | $R^2$ Russia |
|---|---|---|---|---|---|
| P1 | Germany | 0.895 | 0.873 | 0.858 | 0.687 |
| P2 | USA | 0.733 | 0.464 | 0.733 | 0.260 |
| P4 | Germany | 0.722 | 0.197 | 0.373 | 0.161 |
| P5 | Germany | 0.874 | 0.869 | 0.872 | 0.608 |
| P6 | UK | 0.861 | 0.861 | 0.699 | 0.656 |
| P8 | Taiwan | 0.796 | 0.249 | 0.425 | 0.212 |
| P9 | Germany | 0.754 | 0.630 | 0.624 | 0.631 |
| P11 | China | 0.746 | 0.303 | 0.664 | 0.097 |
| P13 | Germany | 0.738 | 0.574 | 0.412 | 0.389 |
| P14 | Germany | 0.708 | 0.510 | 0.546 | 0.087 |
| P20 | USA | 0.912 | 0.787 | 0.912 | 0.774 |
| P21 | SPAIN | 0.791 | 0.620 | 0.727 | 0.720 |
| P22 | USA | 0.870 | 0.176 | 0.870 | 0.111 |
| P23 | USA | 0.874 | 0.659 | 0.874 | 0.517 |
| Global | UK | 0.944 | 0.944 | 0.939 | 0.930 |

**Table 1** – Estimated models for each platform: correlation factors for the countries providing the best fit and for UK, USA and Russia

To summarize, two main findings can be derived from the results presented above:
1) Some trends exhibited at the global level considering the attack processes on all the platforms together are not observed when analyzing each platform individually (this is the case, for example, of attacks from Russia). On the other hand, we have observed the other situation where the trends observed globally are also visible locally on the majority of the platforms (this is the case, for example, of attacks from USA, UK and Germany).
2) The attack processes observed on each platform are very often highly correlated with the attack processes originating from a particular country. The country providing the best regressions locally, does not necessary exhibit high correlations when considering other platforms or at the global level. These trends seem to result from specific factors that govern the attack processes observed on each platform.

### 4.2 Distribution of times between attacks

In this example, we focus on the analysis and the modeling of the times between attacks observed on different honeypot platforms.

Let us denote by $t_i$, the time separating the occurrence of attack $i$ and attack ($i-1$). Each attack is associated to an IP address, and its occurrence time is defined by the time when the first packet is received from the corresponding address at one of the three virtual machines of the honeypot platform. All the packets received from the same IP address within 24 hours are supposed to belong to the same attack session.

We have analyzed the distribution of the times between attacks observed on each honeypot platform. Our objective was to find analytical models that faithfully reflect the empirical data collected from each platform. In the following, we summarize the results obtained considering 5 platforms for which we have observed the highest attack activity.

#### 4.2.1 Empirical analyses

Table 2 gives the number of intervals of times between attacks observed at each platform considered in the analysis as well as the corresponding number of IP addresses. As illustrated by Figure 2, most of these addresses have been observed only once at a given platform. Nevertheless, some IP addresses have been observed several times, the maximum number of visits per IP address for the five platforms was 57, 96, 148, 183, and 83 (respectively). Indeed, the curves plotting the number of IP addresses as a function of the number of attacks for each address follow a heavy-tailed power law distribution. It is noteworthy that such distributions have been observed in many performance and dependability related studies in the context of the Internet, e.g., transfer and interarrival times, burst sizes, sizes of files transferred over the web, error rates in web servers, etc.

|  | P5 | P6 | P9 | P20 | P23 |
|---|---|---|---|---|---|
| Number of ti | 85890 | 148942 | 46268 | 224917 | 51580 |
| Number of IP addresses | 79549 | 90620 | 42230 | 162156 | 47859 |

**Table 2** - Numbers of intervals of times between attacks ($t_i$) and of different IP addresses observed at each platform

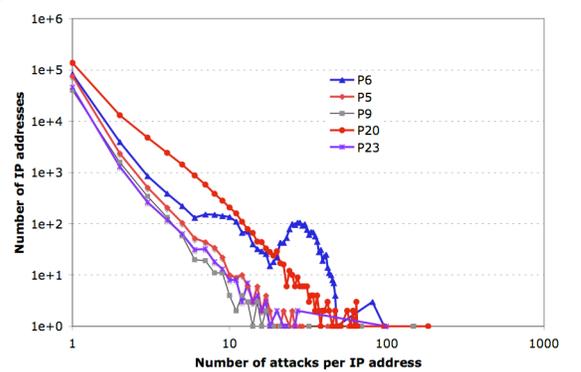

**Figure 2-** Number of IP addresses versus the number of attacks per IP address observed at each platform (log-log scale)

#### 4.2.2 Modeling

Finding tractable analytical models that faithfully reflect the observed times between attacks is useful to characterize the observed attack processes and to find appropriate indicators that can be used for prediction purposes. We have investigated several candidate distributions, including Weibull, Lognormal, Pareto, and the Exponential distribution, which are traditionally used in reliability related studies. The best fit for each platform has been obtained using a mixture model combining a Pareto and an exponential distribution.

Let us denote by *T* the random variable corresponding to the time between the occurrence of two consecutive attacks at a given platform, and *t* a realization of *T*. Assuming that the probability density function *pdf(t)* associated to *T* is characterized by a mixture distribution combining a Pareto distribution and an exponential distribution, then f(t) is defined as follows.

$$pdf(t) = P_a \frac{k}{(t+1)^{k+1}} + (1-P_a)\lambda e^{-\lambda t}$$

$k$ is the index parameter of the Pareto distribution, $\lambda$ is the rate associated to the exponential distribution and $P_a$ is a probability.

We have used the R statistical package [11] to estimate the parameters that provide the best fit to the collected data. The quality of fit is assessed by applying the Kolmogorov-Smirnov statistical test. The results are presented in Figure 3. It can be noticed that for all the platforms, the mixed distribution provides a good fit to the observed data whereas the exponential distribution is not suitable to describe the observed attack processes. Thus, the traditional assumption considered in hardware reliability evaluation studies assuming that failures occur according to a Poisson process does not seem to be satisfactory when considering the data observed form our honeypots. These results have been also confirmed when considering the data collected during other observation periods.

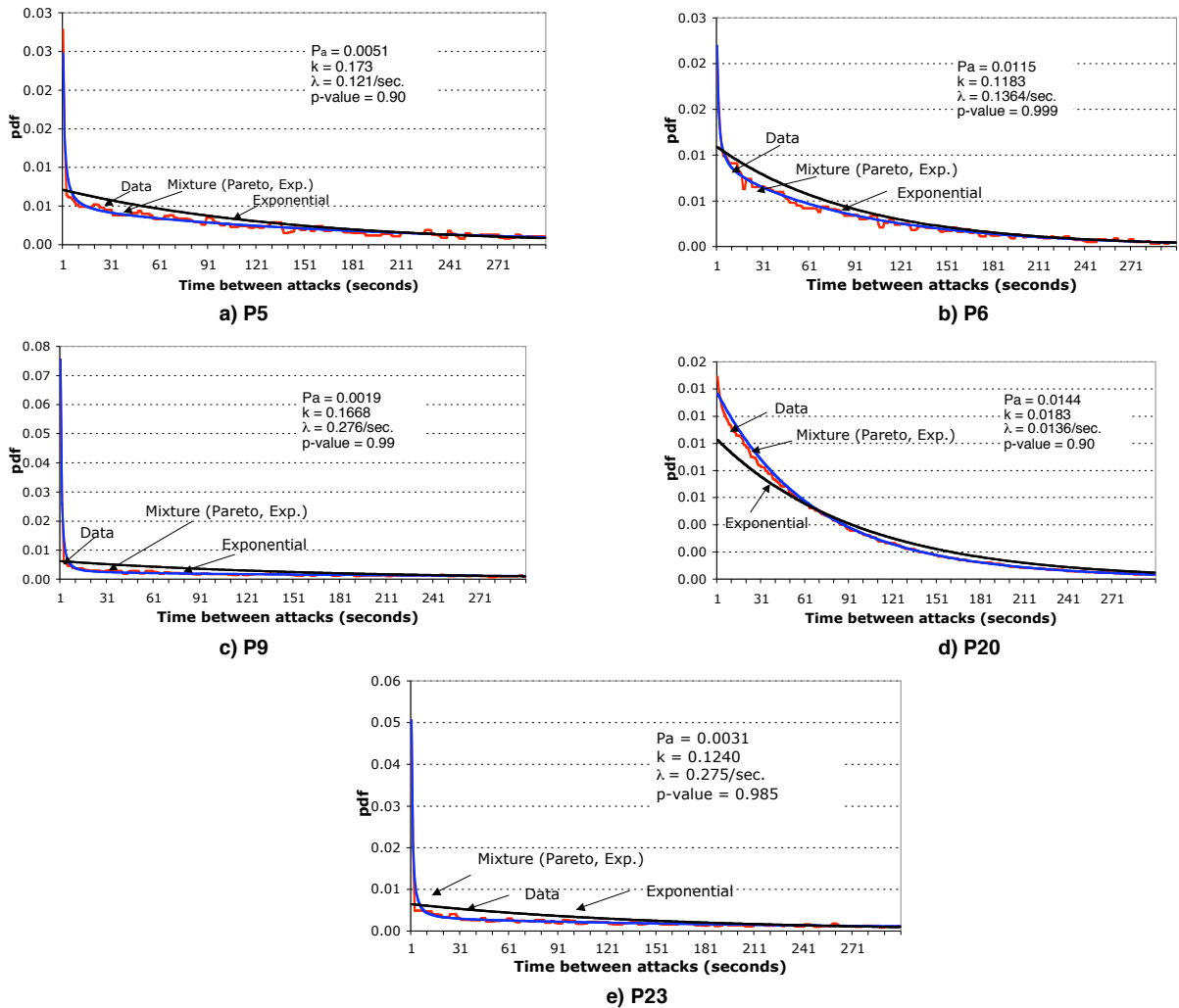

**Figure 3**- Observed and estimated times between attacks probability density functions.

*4.3 Propagation of attacks*

Besides analyzing the attack activities observed at each platform in isolation, it is useful to identify phenomena that reflect propagation of attacks through different platforms. In this section, we analyze simple scenarios where a propagation between two platforms is assumed to occur when the IP address of an attacking machine observed at a given platform is also observed at another platform. Such a situation might occur for example as a result of a scanning activity or might be resulting from the propagation of worms.

For the sake of illustration, we restrict the analysis to the five platforms considered in the previous example. For each attacking IP address in the data collected from the five platforms during the period of the study, we identified: 1) all the occurrences with the same source address, 2) the times of each occurrence and 3) the platform on which each occurrence has been reported. A *propagation* is said to occur for this IP address from platform $P_i$ to platform $P_j$ when the next occurrence of this address is observed on $P_j$ after visiting $P_i$.

Based on this information we build a propagation graph where each node identifies a platform and a transition between two nodes identifies a propagation between the nodes. A probability is associated to each transition to characterize its likelihood of occurrence.

Figure 4 presents the propagation graph obtained for the five platforms included in the analysis. Considering platforms P6 and P20, it can be seen that only a few IP addresses that attacked these platforms have been observed on the other platforms. The situation is different when considering platforms P5, P9, and P23. In particular, it can be noticed that propagation between P5 and P9 is highly probable. This is related in particular to the fact that the addresses of the corresponding platforms belong to the same /8 network domain. More thorough and detailed analyses are currently carried out based on the propagation graph in order to take into account timing information for the corresponding transitions and also the types of attacks observed, in order to better explain the propagation phenomena illustrated by the graph.

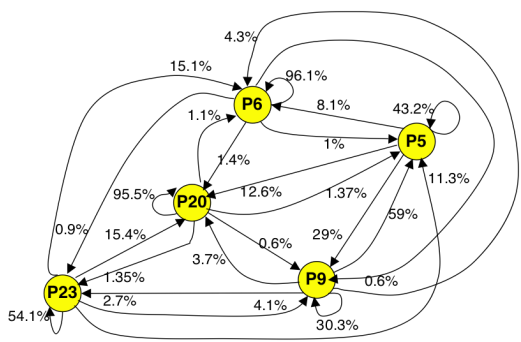

**Figure 4-** Propagation graph

## 5. Conclusion

This paper presented simple examples and preliminary models illustrating various types of empirical analysis and modeling activities that can be carried out based on the data collected from honeypots in order to characterize attack processes. The honeypot platforms deployed so far in our project belong to the family of so-called "low interaction honeypots". Thus, hackers can only scan ports and send requests to fake servers without ever succeeding in taking control over them. In our project, we are also interested in running experiments with "high interaction" honeypots where attackers can really compromise the targets. Such honeypots are suitable to collect data that would enable us to study the behaviors of attackers once they have managed to get access to a target and try to progress in the intrusion process to get additional privileges. Future work will be focused on the deployment of such honeypots and the exploitation of the collected data to better characterize attack scenarios and analyze their impact on the security of the target systems. The ultimate objective would be to build representative stochastic models that will enable us to evaluate the ability of computing systems to resist to attacks and to validate them based on real attack data.

**Acknowledgement.** This work has been carried out in the context of the CADHo project, an ongoing research action funded by the French ACI "Securité & Informatique" (www.cadho.org). It is partially supported by the ReSIST European Network of Excellence (www .resist-noe.org).